\documentclass[journal=apchd5,manuscript=article]{achemso}

\usepackage[version=3]{mhchem} 
\usepackage{amsmath,amssymb,mathrsfs,accents}
\usepackage{color}
\usepackage[scr=rsfso]{mathalpha}
\usepackage{booktabs}                   
\usepackage{multirow}                   
\usepackage{diagbox}                    
\usepackage{bm}
\usepackage{physics}
\usepackage[T1]{fontenc}
\usepackage[utf8x]{inputenc}
\usepackage{filecontents}

\author{Yu-Chen Wei}
\affiliation{Department of Applied Physics and Science Education, Eindhoven University of Technology, 5600MB Eindhoven, The Netherlands}
\email{y.c.wei@tue.nl}
\author{Chih-Hsing Wang}
\affiliation{Department of Chemistry, National Taiwan University, 106319 Taipei, Taiwan}
\author{Konstantinos S. Daskalakis}
\affiliation{Department of Materials Engineering, University of Turku, FI-20014 Turun yliopisto, Finland}
\author{Pi-Tai Chou}
\affiliation{Department of Chemistry, National Taiwan University, 106319 Taipei, Taiwan}
\author{Shunsuke Murai}
\affiliation{Department of Material Chemistry, Graduate School of Engineering, Kyoto University, Kyoto, 615-8510, Japan}
\author{Jaime G\'omez Rivas}
\email{j.gomez.rivas@tue.nl}
\affiliation{Department of Applied Physics and Science Education, Eindhoven University of Technology, 5600MB Eindhoven, The Netherlands}

\title[An \textsf{achemso} demo]
  {Enhanced Delayed Fluorescence in Non-Local Metasurfaces: The Role of Electronic Strong Coupling}

\begin{document}
\begin{tocentry}

    \includegraphics[scale=0.25]{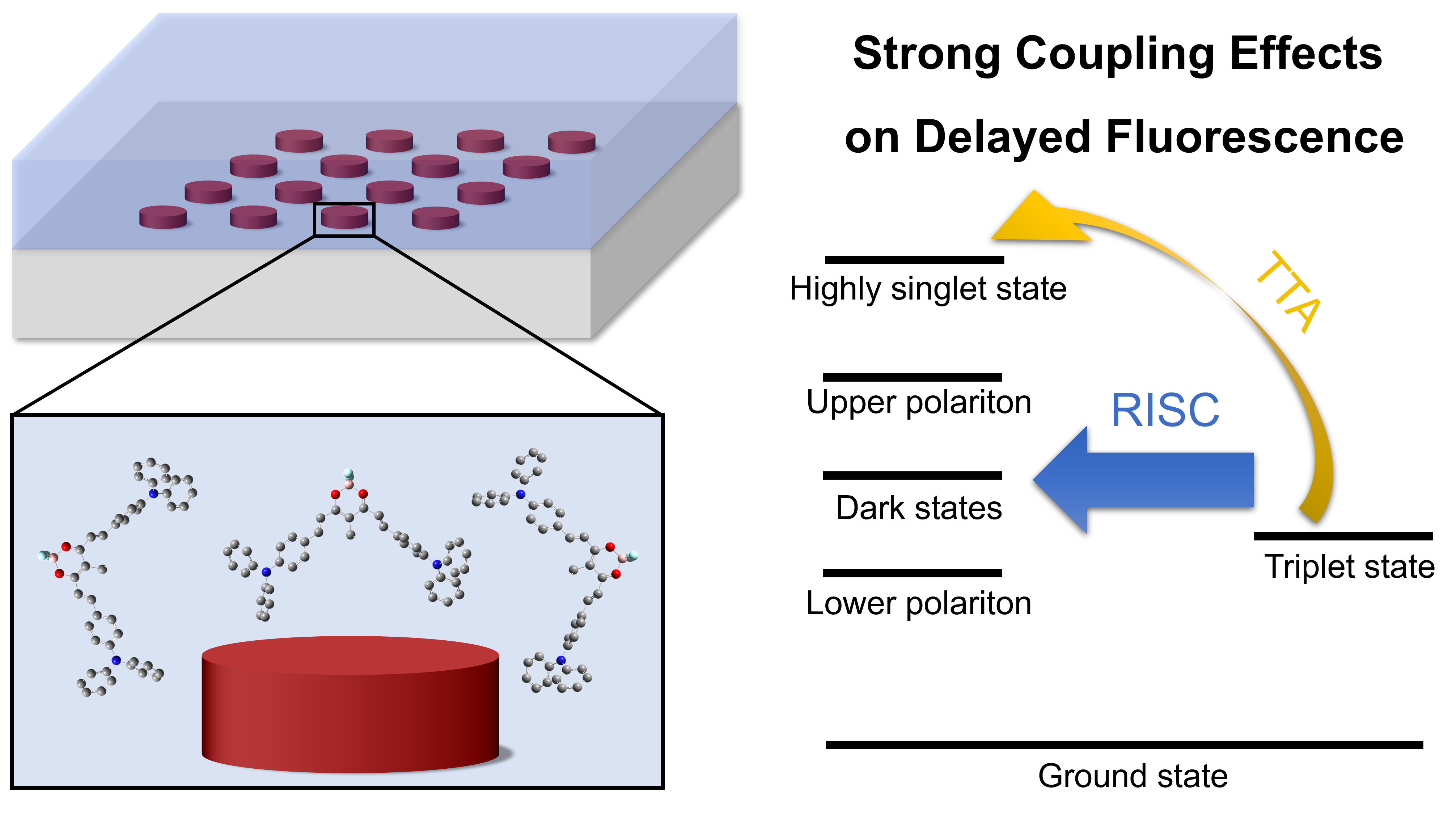}

\end{tocentry}


\begin{abstract}
  Strong light-matter coupling has garnered significant attention for its potential to optimize optoelectronic responses. In this study, we designed open cavities featuring non-local metasurfaces composed of aluminum nanoparticle arrays. The surface lattice resonances in these metasurfaces exhibit electronic strong coupling with the boron difluoride curcuminoid derivative, known for its highly efficient thermally activated delayed fluorescence in the near-infrared. Our results show that delayed fluorescence induced by triplet-triplet annihilation can be enhanced by a factor of 2.0-2.6 in metasurfaces that are either tuned or detuned to the molecular electronic transition. We demonstrate that delayed fluorescence enhancements in these systems primarily stem from increased absorption in the organic layer caused by the nanoparticle array, while strong coupling has negligible effects on reverse intersystem crossing rates, aligning with previous studies. We support these findings with finite-difference-time-domain simulations. This study elucidates how light-matter interactions affect delayed fluorescence, highlighting the potential applications in optoelectronic devices.
\end{abstract}

Thermally activated delayed fluorescence (TADF) organic emitters have demonstrated the ability to achieve 100\% internal quantum efficiency (IQE) in organic light emitting diodes (OLEDs) through an efficient reverse intersystem crossing (RISC) mechanism \cite{uoyama2012highly,wong2017purely}. This is particularly significant because the molecular design of TADF emitters achieves such high IQE without the need for heavy metals, unlike phosphorescent emitters \cite{chou2007phosphorescent,yang2015functionalization}. Instead, this efficiency is realized by designing compounds with a sub-100~meV energy gap between the lowest singlet excited (S$_1$) state and the lowest triplet (T$_1$) state, exhibiting mixed localized excited (LE) and charge-transfer (CT) characters \cite{etherington2016revealing}. In addition to TADF dyes, some high-efficiency OLEDs,  especially deep blue OLEDs, have employed triplet-triplet annihilation (TTA) emitters to convert dark triplet excitons into bright singlet excitons \cite{jiang2023recent,gao2021application}. TTA arises when the population of triplet excitons is sufficient. The annihilation between two triplet excitons generates a spin-correlated complex. Under conservation of the total spin and energy, the spin-correlated complex would dissociate into one singlet exciton with the energy of the two triplet states, whereas the other one returns to its ground state \cite{xiao2022situ}.

A current challenge, however, is that RISC rates in TADF emitters are typically below 10$^6~\mathrm{s^{-1}}$, and efforts to push these rates to higher values often result in reduced singlet oscillator strengths, thereby lowering singlet radiative rates and ultimately limiting the brightness \cite{song2023regulating,kim2020nanosecond}. In the case of TTA emitters, achieving high efficiency is hindered by efficiency roll-off at high brightness levels and the energy alignment requirements between one singlet exciton and two triplet excitons, which complicate the development of highly efficient TTA emitters \cite{reineke2007triplet,zhang2012triplets}. Conventionally, efforts to address this challenge have focused on developing new TADF emitters\cite{uoyama2012highly,wong2017purely} or TTA emitters\cite{jiang2023recent,gao2021application}, often involving detailed molecular simulations and material synthesis. While these approaches have significantly contributed to the field, the growing demand of OLEDs highlights the importance of exploring alternative strategies. Therefore, a more efficient approach to optimize emission is needed to accelerate the development of OLEDs. 

Recent studies have shown that strong light-matter coupling can alter excited-state dynamics \cite{virgili2011ultrafast,hutchison2012modifying,schwartz2013polariton,wang2014quantum,balasubrahmaniyam2023enhanced,tibben2023molecular,zeng2023control}, thereby influencing optoelectronic performance \cite{nikolis2019strong,tang2024strong,de2024enhancement,mischok2023highly,witt2024high}. Strong light-matter coupling leads to the formation of light-matter hybrid states, the so-called polariton states. These hybrid states are referred to as the lower (LP) and upper polaritons (UP). The energy difference between the LP and UP, also known as the Rabi splitting, indicates the strength of the light-matter coupling \cite{ebbesen2016hybrid,bhuyan2023rise}. To achieve strong light-matter coupling, it is necessary to confine optical modes via specific photonic environments, such as photonic or plasmonic cavities \cite{simpkins2023control}. These cavities can modify the emission characteristics without the need of developing new TADF dyes, and their optical responses are predictable via electrodynamics simulations, which accelerate OLED development \cite{mischok2023highly,witt2024high}.

Among various photonic cavities, the Fabry-Perot cavity has been extensively applied to achieve strong light-matter coupling due to its ease of fabrication \cite{bhuyan2023rise,simpkins2023control}. In 2019, Elad {\it et al.} investigated the inverted singlet-triplet energy gap caused via electronic strong coupling in Fabry-Perot cavities and its effects on RISC processes \cite{eizner2019inverting}. Their findings indicated that strong coupling had negligible effects on TADF dynamics, primarily due to the dominant density of states (DOS) associated with dark states. Inseong {\it et al.} showed that strong coupling reduces prompt and delayed emission through excimer states due to efficient energy transfer to the LP states \cite{cho2023multi}. Abdelmagid {\it et al.} reported the emergence of delayed electroluminescence in a fluorescent polariton OLED, but its origin lies in the enhanced outcoupling
emission of trapped charges in the cavity \cite{abdelmagid2024identifying}. Moreover, Tomohiro {\it et al.} observed a reduction in the prompt and delayed emission rates of a multiresonance TADF dye under strong coupling regime \cite{ishii2024modified}. In addition to TADF, Daniel {\it et al.} demonstrated that electronic strong coupling modified the efficiencies of singlet fission and TTA, thereby increasing delayed fluorescence (DF) \cite{polak2020manipulating}. Beyond Fabry-Perot closed cavities, metasurfaces form open-cavity systems that are suitable candidates to achieve the strong coupling regime while maintaining emission efficiency. Berghuis {\it et al.} have shown that strong coupling between tetracene crystals and surface lattice resonances (SLRs) in metasurfaces enhances the DF yield by a factor of four \cite{berghuis2019enhanced}. However, the detailed mechanism of this enhancement remains unclear due to the complexity of multiple excited-state processes in the tetracene crystals.

\begin{figure}[h!]
    \centering
    \includegraphics[width=\linewidth]{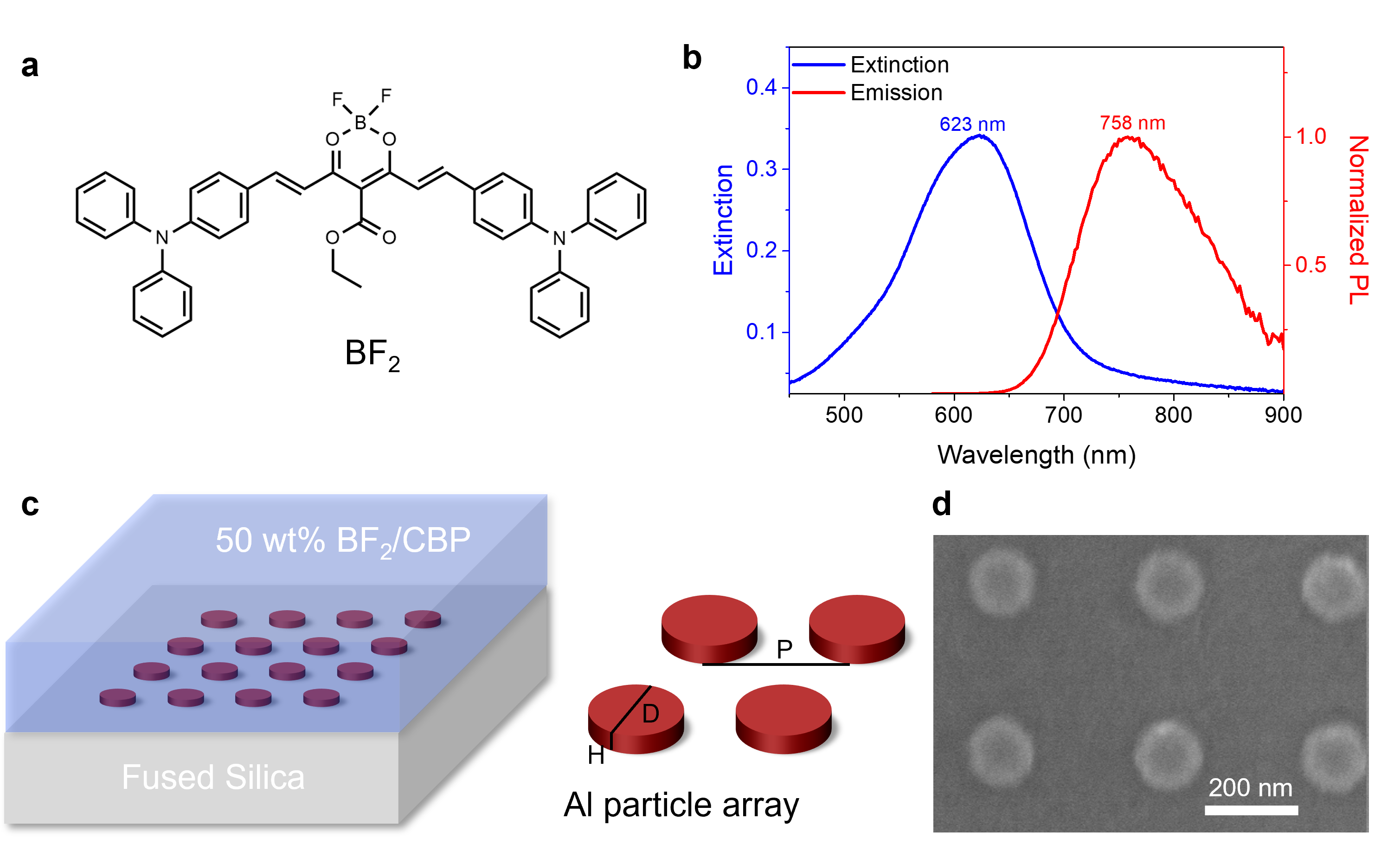}
    \caption{TADF molecule and the metasurfaces. (a). Molecular structure of BF$_2$. (b) Steady-state extinction and emission spectra of 50 wt\% CBP/BF$_2$ blended thin film. The tail in the extinction above 700 nm is caused by reflections at the interface air/film \cite{leppala2024linear}. (c) Illustration of an aluminum nanoparticle array. The thickness of the organic layer is 100 nm. P, D and H indicate the period, the diameter, and the height of the nanoparticle array. (d) SEM image of the array with a SLR tuned to the S$_0$-S$_1$ transition of BF$_2$.}
    \label{fig1}
\end{figure}

In this study, we investigate how strong coupling in metasurfaces influences molecular DF dynamics. For the TADF system, we choose the boron difluoride curcuminoid derivative (BF$_2$) (Figure~\ref{fig1}a), which meets multiple criteria for studying the effects of strong coupling on DF. First, BF$_2$ exhibits a high absorption coefficient for S$_0$-S$_1$ transition ($>10^5$ M$^{-1}$cm$^{-1}$.) \cite{kim2018high}, which is essential to achieve the strong coupling regime. Second, BF$_2$ has TADF even at high doping concentrations (over 50 \%) \cite{kim2018high}, which is rare for TADF dyes, as high concentrations typically quench triplet excitons and reduce TADF yield \cite{adachi2001endothermic,dias2013triplet}. Third, unlike tetracene crystals, the excited-state processes of BF$_2$ do not involve singlet fission or triplet dissociation, simplifying its analysis \cite{berghuis2019enhanced}. The photophysical properties of the BF$_2$ are characterized in the thin film with 50$\%$ blended in a 4,4'-bis(N-carbazolyl)-1,10-biphenyl (CBP) host. A 50$\%$ concentration is selected to achieve strong coupling while preserving good film quality, high DF yields, and emission intensity. The absorption and emission spectra of the blended film are presented in Figure~\ref{fig1}b. The refractive index of the film is characterized by ellipsometry (Figure~S1), which shows a high extinction coefficient $k = 0.54$ monitored at the wavelength of the absorption peak (623 nm). The high $k$ value indicates the large transition dipole moment of BF$_2$, facilitating strong light-matter interaction.

To design the metasurfaces for electronic strong coupling, we simulate the transmission spectra of the blended film on metasurfaces using the finite-difference-time-domain (FDTD) method \cite{FDTD}. The simulated structure consists of a fused silica substrate and the aluminum (Al) nanoparticle array embedded within the 50$\%$ BF$_2$/CBP blended film (Figure~\ref{fig1}c). The SLRs, resulting from the enhanced radiative coupling of the nanoparticles through in-plane diffraction by the array, are simulated with the arrays embedded in a dielectric material (thickness= 100 nm) with a constant real refractive index ($n=2.0$), and zero material dissipation ($k = 0$) according to the averaged $n$ value in the BF$_2$/CBP blended film (Figure~S1). For these simulations, a hypothetical non-dissipative medium is used to numerically characterize the SLRs in the absence of molecular electronic transitions. By assigning a real refractive index without absorption to the thin film, we can tune the SLRs at normal incidence to match the frequency of the molecular transition responsible for light-matter coupling (Figure~S2) \cite{verdelli2024polaritonic,verdelli2024ultrastrong}. The optimized geometric parameters for the tuned arrays include Al disks with diameter D = 120 nm, height H = 30 nm, and period P = 350 nm arranged in a square array. To investigate the DF dynamics on metasurfaces in different coupling regimes, we also design blueshifted detuned (B-detuned) and redshifted-detuned (R-detuned) metasurfaces with the same disk diameters and heights but different periods (P = 300 nm for B-detuned and P = 400 nm for R-detuned). 

According to the structural parameters from the simulation, we fabricated the particle array using electron beam lithography (EBL) on a fused silica substrate (thickness = 0.5 mm). Details of the sample fabrication EBL are given in Materials and Methods. The dimensions of the arrays are 7.5 mm $\times$ 7.5 mm. A scanning electron microscope (SEM) image of a portion of the tuned array (P = 350 nm) is shown in Fig.~\ref{fig1}d. The SEM images of the B-detuned and R-detuned arrays are shown in Figure~S3.

\begin{figure}[h!]
    \centering
    \includegraphics[width=\linewidth]{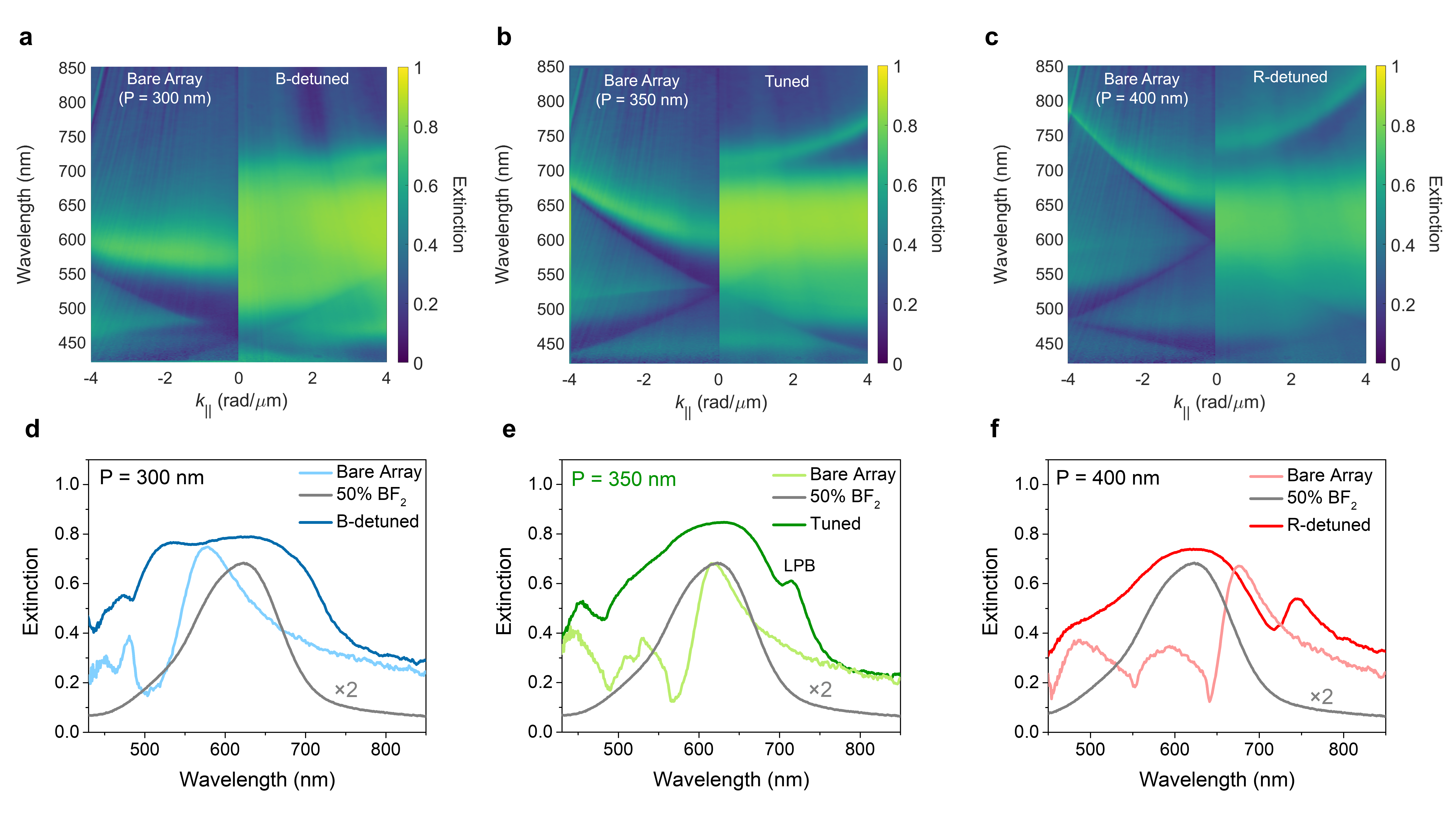}
    \caption{Angle-resolved optical extinction measurements of the tuned and detuned systems with TE illumination. Extinction maps of (a) the B-detuned system (P = 300 nm); (b) the tuned system (P = 350 nm), (c) the R-detuned system (P = 400 nm). The left parts of (a)-(c) correspond to the extinction maps of the bare arrays (CBP thin films on arrays). The right parts of (a)-(c) correspond to the extinction maps of 50 wt\% CBP/BF$_2$ blended thin film on the tuned and detuned metasurfaces. Extinction spectra of (d) the B-detuned system, (e) the tuned system, and (f) the R-detuned system. All spectra are taken at $k_\parallel$ = 1 rad/$\mu$m. The gray lines in (d), (e) and (f) represent the emission spectra of 50 wt\% CBP/BF$_2$ blended thin film (non-cavity system).}
    \label{fig2}
\end{figure}  

The light-matter coupling between BF$_2$ and SLRs is characterized by the angle-resolved extinction measurements (Figure~\ref{fig2}). These measurements were performed with a Fourier microscope under transverse electric (TE) excitation (See details in Materials and Methods). The extinction is analyzed as a function of wavelength and the in-plane wavevector of the incident beam parallel to the surface ($k_\parallel$). To characterize the mode dispersion of the bare arrays, i.e. the SLRs without BF$_2$ molecules, we spincoated a neat CBP film on top of the metasurfaces as an index-matching material (n = 1.9) \cite{liu2005characterization,morozov2019revising}. The extinction maps of the bare arrays reveal parabolic-shaped bands with increased extinction above 550 nm, corresponding to SLRs (the left parts of Figure~\ref{fig2}a, \ref{fig2}b and \ref{fig2}c). Next, we prepared the tuned and detuned systems by replacing the CBP layer with the 50 wt\% CBP/BF$_2$ thin film. Their extinction maps are shown in the right parts of Figure~\ref{fig2}a, \ref{fig2}b and \ref{fig2}c. In the tuned system, a significant wavelength shift of the parabolic band indicates the formation of electronic strong coupling (Figure~\ref{fig2}b), corresponding to the lower polaritonic band (LPB).

To better visualize the polaritonic bands, extinction spectra are plotted for the 50 wt\% CBP/BF$_2$ thin film, the bare arrays, and the tuned and detuned systems measured at $k_\parallel$ = 1 rad/$\mu$m (Figure~\ref{fig2}d, \ref{fig2}e and \ref{fig2}f). In the tuned system, the SLR peak aligns with the molecular absorption peak, facilitating the formation of electronic strong coupling. In addition, the LPB arises at 715 nm while the upper polaritonic band (UPB) is indistinguishable due to the broad molecular absorption peak and interference from high-order modes (Figure~\ref{fig2}e). In contrast, the B-detuned and R-detuned systems exhibit shifted SLRs modes, attributed to the refractive index differences between CBP and 50 wt\% CBP/BF$_2$ within the absorption region of BF$_2$ (Figure~\ref{fig2}d and \ref{fig2}f). These results indicate that SLRs in the B-detuned and R-detuned systems are not as strongly coupled to molecules as in the tuned system. Similar results were obtained under transverse magnetic (TM) excitation (Figure~S4).

To quantify the light-matter coupling strength, we fit the angle dispersion of the LPB to the coupled harmonic oscillator model
\begin{align}
\label{Eq:H}
H=\begin{pmatrix}
E_\mathrm{SLR}-i\gamma_\mathrm{c} & g \\
g & E_\mathrm{exc}-i\gamma_\mathrm{m}
\end{pmatrix}\;,
\end{align}
where $E_\mathrm{SLR}$ is the wave-vector-dependent energy of the SLRs, $E_\mathrm{exc}$ is
the energy of the 50 wt\% CBP/BF$_2$ $\mathrm{S}_1$ exciton (1.99 eV), $\gamma_\mathrm{c}$ and $\gamma_\mathrm{m}$
are the losses of the photonic modes and the molecular transition, respectively, estimated from the full width at half maxima of the extinction spectra, and $g$ is the light-matter coupling strength. $E_\mathrm{SLR}$ is determined by fitting the left part of Figure~\ref{fig2}b using a model that accounts for the coupling between the localized surface plasmon resonances in the individual nanoparticles and the in-plane diffraction order or Rayleigh anomaly (details provided in the Materials and Methods section). The fits are shown in Figure~S5. The fitting parameters are $g= 220~\mathrm{meV}$, $\gamma_\mathrm{c}= 23~\mathrm{meV}$ and $\gamma_\mathrm{m}= 200~\mathrm{meV}$. According to the Savona {\it et al.} criterion for strong coupling: $2g > |\gamma_c-\gamma_m|$ \cite{savona1995quantum}, the tuned system is in the strong coupling regime. 

\begin{figure}[h!]
    \centering
    \includegraphics[width=\linewidth]{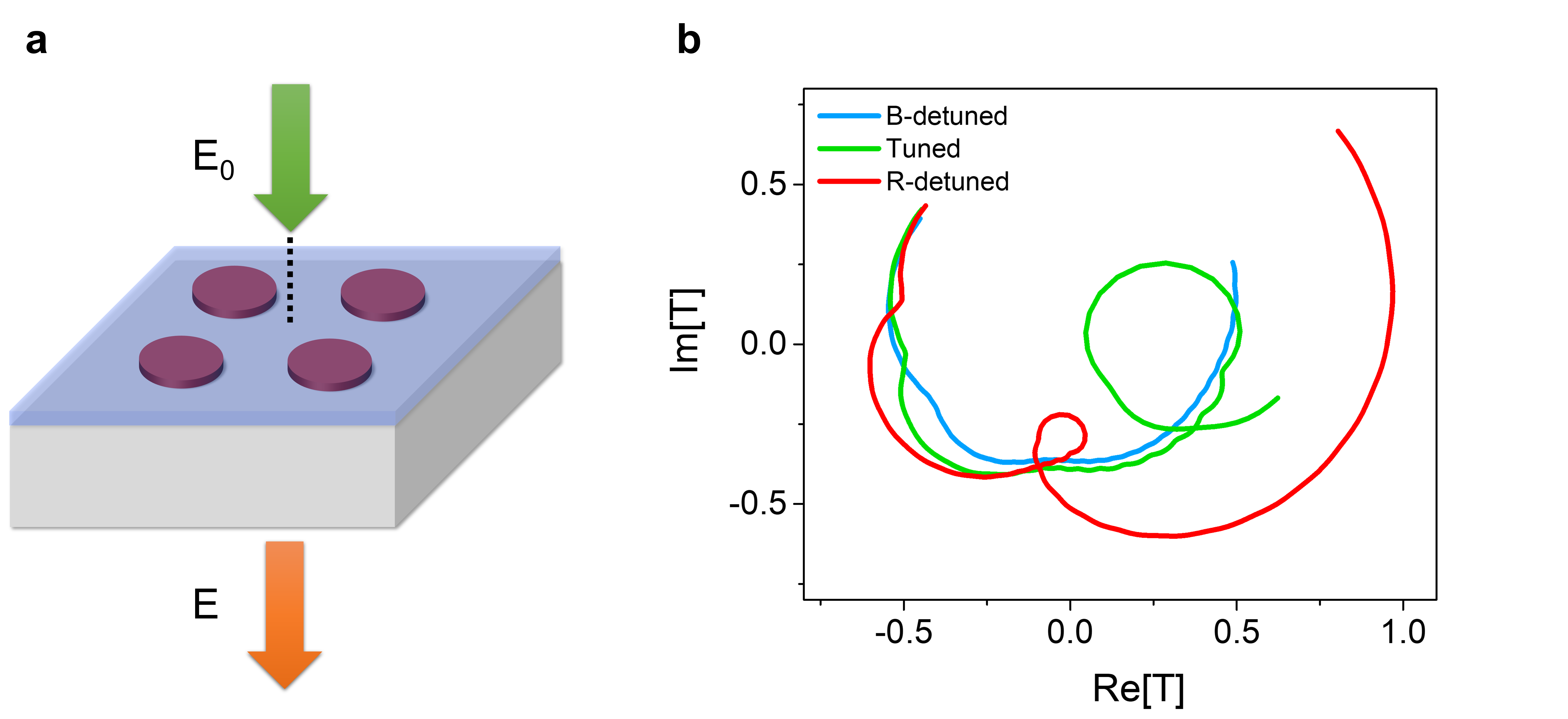}
    \caption{Complex transmittance simulations. (a) Schematic of the simulation, where E and E$_0$ are the transmitted electric field and the source, respectively. The complex transmittance T is defined as $\mathrm{T}=\mathrm{E}/|\mathrm{E}_0|$. (b) Complex transmittance simulations of the tuned system, and the B-detuned and the R-detuned systems.}
    \label{fig3}
\end{figure}

To further confirm the realization of electronic strong coupling, we analyze the existence of a secondary loop in the complex transmittance plane (Figure~\ref{fig3}). The change in phase topology from weak to strong coupling has been introduced recently by Thomas {\it et al.} as an alternative method to probe strong coupling \cite{thomas2020new}. We applied the FDTD method to simulate the complex transmittance T at normal incidence ($k_\parallel$ = 0 rad/$\mu$m). The complex T is defined by $\mathrm{T}=\mathrm{E}/|\mathrm{E}_0|$, where E and E$_0$ are the source and transmitted electric field, respectively. The results, displayed in Figure~\ref{fig3}, show a loop in the complex transmittance plane for the tuned system and the R-detuned system, indicating electronic strong coupling, and a different topology (no loop) for the B-detuned system, indicating weak coupling for this system.

\begin{figure}[h!]
    \centering
    \includegraphics[width=\linewidth]{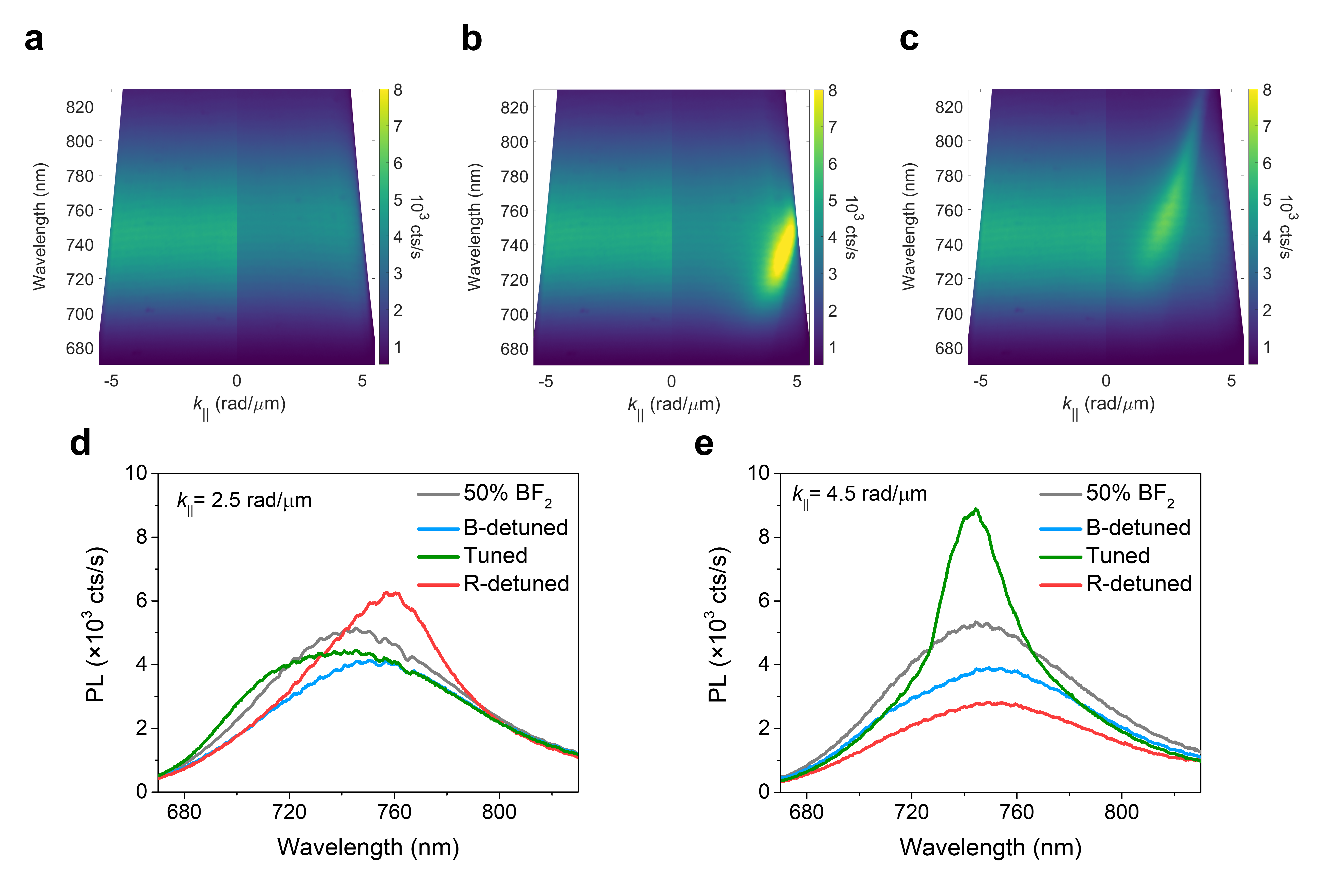}
    \caption{TE-polarized angle-resolved optical PL measurements of the tuned and detuned systems. Steady-state PL maps of the (a) B-detuned, (b) tuned, and (c) R-detuned systems. The left parts of (a)-(c) correspond to the PL map of the non-cavity system (bare 50 wt\% CBP/BF$_2$ thin film). PL spectra of the tuned/detuned and non-cavity systems measured at (d) $k_\parallel$ = 2.5 rad/$\mu$m and (e) $k_\parallel$ = 4.5 rad/$\mu$m, respectively.}
    \label{fig4}
\end{figure}

The steady-state TE-polarized and angle-resolved photoluminescence (PL) of the tuned and detuned systems are characterized in Figure~\ref{fig4}.  In Figure~\ref{fig4}b, the tuned system exhibits a significant PL enhancement of 1.7-fold compared to the non-cavity system at $k_\parallel$> 4 rad/$\mu$m. 
Note that the PL enhancement occurs only at specific wave vectors. For the tuned system, the PL spectra at $k_\parallel=4.5$ rad/$\mu$m show the enhancement corresponding to the LPB, whereas the PL intensity decreases at $k_\parallel=2.5$ rad/$\mu$m (Figures~\ref{fig4}d and \ref{fig4}e). In contrast, the PL of the B-detuned system is weaker than the non-cavity system for all $k_\parallel$ due to the emission quenching from the Al nanoparticles (Figure~\ref{fig4}a). For the R-detuned system, slight PL enhancement is observed due to the detuned SLR mode coupled to the 50 wt\% CBP/BF$_2$ film (Figure~\ref{fig4}c and \ref{fig4}d), though the enhancement is smaller than that associated with the LPB of the tuned system. Similar phenomena are also visible for the TM-polarized emission (Figure~S6). Importantly, the total PL intensity, averaged over -5 rad/$\mu$m < $k_\parallel$ < 5 rad/$\mu$m, is lower for all systems compared to the non-cavity system regardless of polarizations (Figure~S7). This result indicates that while SLRs and electronic strong coupling modify light outcoupling, they do not macroscopically enhance the photoluminescence quantum yield.

\begin{figure}[h!]
    \centering
    \includegraphics[width=\linewidth]{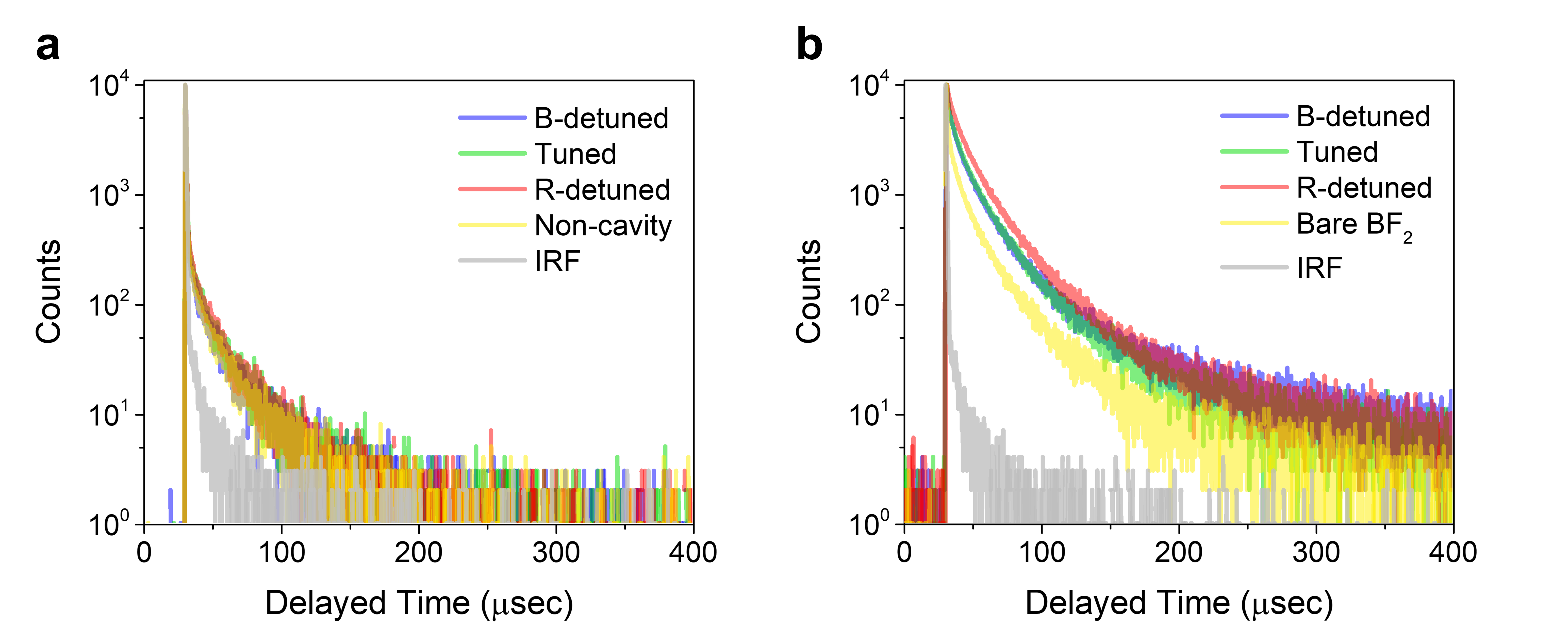}
    \caption{Dynamics of DF in the tuned, detuned, and non-cavity systems. (a). Time-resolved PL under low excitation power density (0.08 mW/mm$^2$).  (b) Time-resolved PL under high excitation power density (10.28 mW/mm$^2$). The excitation and detection wavelengths are 532 nm and 750 nm, respectively. IRF indicates the instrumental response function.}
    \label{fig5}
\end{figure}

To elaborate on the strong coupling effects on the TADF dynamics, we measured the time-resolved PL with a time-correlated single photon counting (TCSPC) system (Figure~\ref{fig5}). Details of the TCSPC setup are provided in Materials and Methods. To ensure that the measured PL originates primarily from TADF and not from other nonlinear effects, we conducted measurements at a low excitation power density of 0.08 mW/mm$^2$. Under this condition, the PL emission time traces reveal both prompt fluorescence and DF (Figure~\ref{fig5}a). According to previous studies, the DF arises from both monomeric and dimeric species \cite{kim2018high}. In addition, the emission time traces on the hundred-microsecond timescale exhibit negligible differences among the tuned, detuned, and non-cavity systems. Similarly, the nanosecond-scale emission time traces also show no significant variation between these systems (Figure~S8). These results indicate that strong coupling and SLRs do not significantly influence the TADF dynamics, which is consistent with previous reports suggesting that the low density of states of the LPB compared to dark states, limits the modification of the PL emission dynamics.\cite{eizner2019inverting,abdelmagid2024identifying,cho2023multi,ishii2024modified}. 

In contrast, as the excitation power density increases, the DF component grows in all the systems due to TTA (Figure~S9). The nonlinear character of the increased DF is validated via the power-dependent PL measurements (Figure~S10). With the rise of TTA, the DF in the tuned and detuned systems surpasses that of the non-cavity system (Figure~\ref{fig5}b). Notably, the highest DF enhancement is observed in the R-detuned system, while the enhancements in the tuned and B-detuned systems are comparable. To rule out artifacts arising from the high-power excitation, we measured the emission time trace of the fused silica substrate, which confirms that the DF under high excitation power is not related to scattering or emission from the substrate (Figure~S11). 

To quantify the DF enhancements, the emission time traces are fitted with a triple exponential function $I(t)=A_1 e^{t/\tau_1}+A_2 e^{t/\tau_2}+A_3 e^{t/\tau_3}$, where $I(t)$ is the PL intensity as a function of time, $A_1 e^{t/\tau_1}$ represents the prompt population decay convoluted with the instrument response function (IRF). $A_2 e^{t/\tau_2}$ and $A_3 e^{t/\tau_3}$ represent the DF caused via the monomeric and the dimeric species, respectively \cite{kim2018high}. The fitted results are shown in Figure~S12. Next, we define the intensity of the DF, $I_\mathrm{DF}$, by integrating $A_2 e^{t/\tau_2}$ and $A_3 e^{t/\tau_3}$ over time from zero to infinity: 
\begin{align}
\label{Eq:I_df}
I_\mathrm{DF}&=\int_0^\infty dt(A_2 e^{t/\tau_2}+A_3 e^{t/\tau_3})\nonumber\\
&=A_2\tau_2+A_3\tau_3 \;.
\end{align}
We define $I_\mathrm{DF,0}$ as the $I_\mathrm{DF}$ of the non-cavity system. Based on this definition, we evaluate the DF enhancements $I_\mathrm{DF}/I_\mathrm{DF,0}$ for the tuned/detuned systems (Table 1). Under low-power excitation, the DF enhancements of all systems are close to 1. However, under high-power excitation, the tuned and detuned systems exhibit DF enhancements exceeding 2, with the R-detuned system showing the highest enhancement of 2.61-fold. These results indicate that the observed DF enhancements are attributed to the metasurfaces rather than electronic strong coupling.

\renewcommand{\arraystretch}{2}
\begin{table}[!h]
\caption{Comparison of DF enhancements. 
The error bars result from the deviation between the measured value and the fitted curve.}
\begin{tabular}{c|c|c|c} 
{Excitation Power Density (mW/mm$^2$)} & B-detuned & Tuned & R-detuned \\
\hline
{0.08} & 1.09 $\pm$ 0.09 & 1.13 $\pm$ 0.10 & 0.83 $\pm$ 0.07 \\
\hline
{10.28} & 2.04 $\pm$ 0.04 & 2.14 $\pm$ 0.04 & 2.61 $\pm$ 0.06  \\
\end{tabular}
\label{Table:1}
\end{table}
\renewcommand{\arraystretch}{1}

\begin{figure}[h!]
    \centering
    \includegraphics[width=\linewidth]{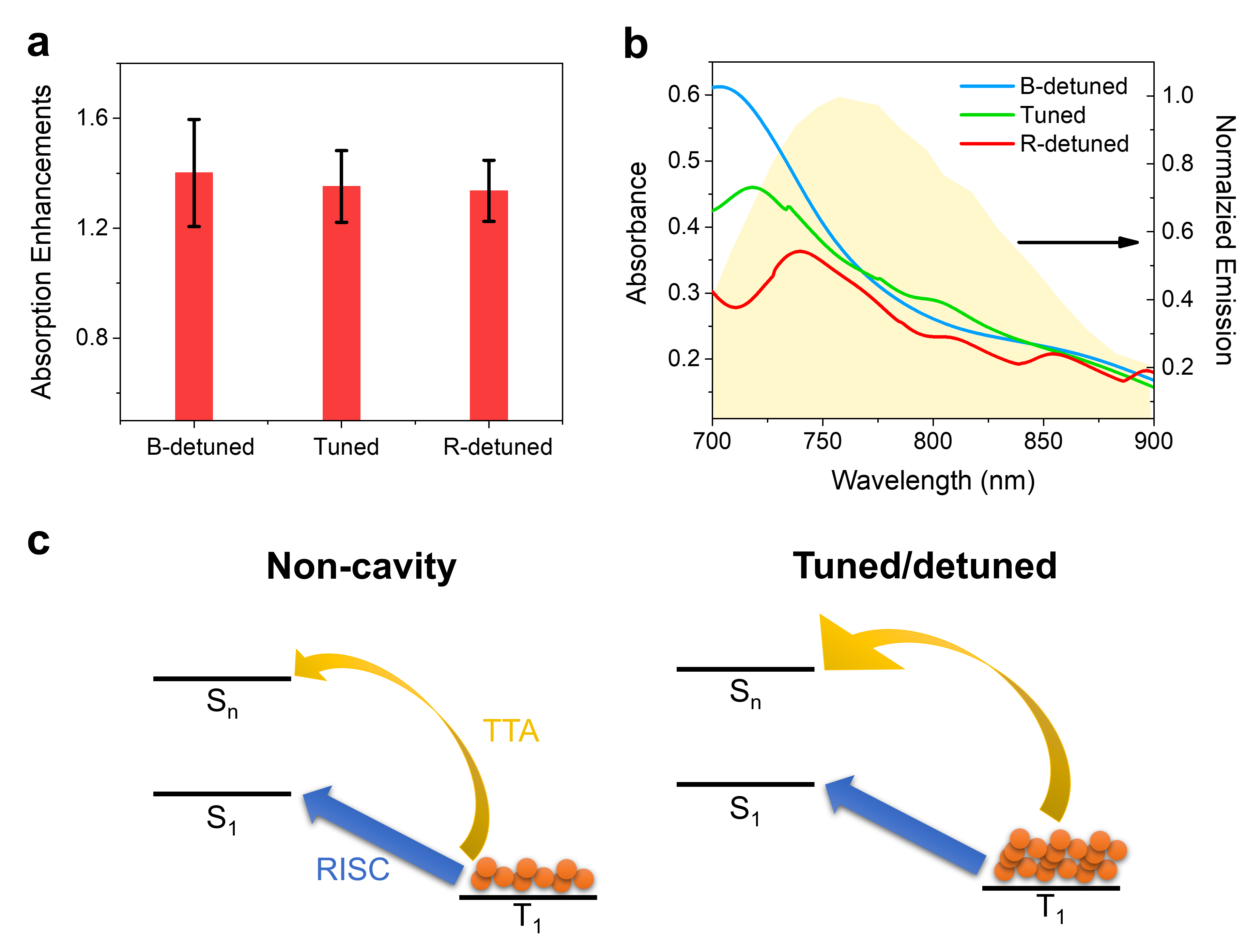}
    \caption{Mechanism of TTA in the tuned and detuned systems. (a) Simulation of absorbed power at the excitation wavelength. The absorption enhancements are obtained from the absorbed power in the organic layer in the tuned/detuned systems divided by the absorbed power in the non-cavity system. The error bars arise from the different incident angles due to the focused excitation beam. (b) Simulation of the absorbance spectra by the whole systems (organic layers and nanoparticle arrays) in the wavelength range of emission. The light-yellow area indicates the emission profile of the non-cavity system. The overlaps between the absorbance spectra of the emission spectra provides the emission quenching. (c) Schematic illustration of DF dynamics in the different systems. RISC indicates the reverse intersystem crossing process. TTA represents the triplet-triplet annihilation process. The sizes of the arrows represent the transition rates.}
    \label{fig6}
\end{figure}

 To explain the DF enhancements observed in both tuned and detuned systems, we propose that the electromagnetic field absorption caused by the nanoparticle arrays increases the triplet population, thereby enhancing the TTA-induced DF. To test this hypothesis, we used the FDTD method to simulate the absorption enhancements of the organic layers at the excitation wavelength of 532 nm (Figure~\ref{fig6}a). Here, the absorption enhancement is defined as the ratio of power absorbed in the organic layer in the tuned/detuned systems to that in the non-cavity system. Importantly, the absorption in the Al nanoparticles was excluded in the simulations, focusing solely on the organic layer (see Materials and Methods). The simulation follows the experimental setup of the TCSPC, with the incident polar angle set to 45$^\circ$ to minimize the specular reflection of the excitation pulses. Due to beam focusing, the incident azimuthal and polar angle ranges from 25$^\circ$ degree to 35$^\circ$ and 40$^\circ$ degree to 50$^\circ$, respectively. These introduce small variations in the calculated absorption enhancements. Figure~\ref{fig6}a shows that the tuned and detuned arrays exhibit absorption enhancements greater than 1 (ranging from 1.34 to 1.40). These results support our hypothesis that photonic modes on the metasurfaces enhance the absorption within the organic layer, regardless of whether electronic strong coupling is formed. This increased absorption leads to a higher triplet population and, subsequently, greater TTA-induced DF compared to the non-cavity system. 

Although Figure~\ref{fig6}a explains why the DF on metasurfaces is more pronounced than in the non-cavity system, the slight differences in absorption enhancements among the tuned and detuned systems are inconsistent with the trend of DF enhancements (Table 1). This discrepancy suggests that an additional factor influences DF dynamics. This factor is the emission quenching from the Al nanoparticle arrays. As shown in Figure~\ref{fig6}b, the spectral overlap between the simulated absorbance spectra of the tuned/detuned systems, which includes the material dissipation from both the organic layers and the nanoparticle arrays, and the experimental emission spectrum of the non-cavity system is the largest for the B-detuned system, while the R-detuned system has the smallest spectral overlap. This trend arises from the varying densities of nanoparticles. The smaller spectral overlap in the R-detuned system suggests that its PL is less susceptible to quenching by the nanoparticle array, which accounts for its highest DF enhancement. 

Based on the above results, we summarize the electronic strong coupling effects on the DF dynamics of the 50 wt\% CBP/BF$_2$ blended thin film in Figure~\ref{fig6}c. In the non-cavity system, DF arises from reverse intersystem crossing (RISC, an intramolecular process) and TTA (an intermolecular process) under high excitation power. For the RISC pathway, neither electronic strong coupling nor detuned SLRs affect its dynamics, as validated by the time-resolved PL measurements under low excitation power (Figure~\ref{fig6}a). This finding is consistent with previous reports indicating that the low DOS of polaritonic modes has negligible effects on RISC, with most population transitions occurring for the dark states or uncoupled states \cite{eizner2019inverting,abdelmagid2024identifying}. In contrast, for the TTA process, the tuned and the detuned systems exhibit higher TTA-induced DF due to the absorption enhancements in the organic layer caused by the nanoparticle arrays.

Our key findings are summarized as follows. First, we demonstrate that the electronic transition of the BF$_2$ can couple with the SLRs in Al metasurfaces, achieving the electronic strong coupling regime. Second, the time-resolved PL shows that the existence of electronic strong coupling do not affect the RISC processes, leading to negligible changes in TADF dynamics. Third, the absorption enhancements at the excitation wavelength increase the TTA-induced DF by a factor of 2.0-2.6, regardless of the formation of electronic strong coupling. Fourth, the trend of the DF enhancement is governed by two factors: increased absorption at the excitation wavelength, and PL quenching caused by absorption in the nanoparticle arrays. This study provides new insights in the role of electronic strong coupling and optical modes in metasurfaces for the modified DF dynamics of TADF molecules. It will help in further designs of metasurfaces for improved light emission by emphasizing on the relevant mechanisms for these improvements.

\section*{Materials and Methods}
\subsection*{FDTD Simulation}
We used Lumerical FDTD Solutions to perform the simulations. The complex permittivity of Al from Fei {\it et al.}\cite{cheng2016epitaxial} and of the BF$_2$ thin film from the ellipsometry measurements were used for these simulations. The substrate is set as a non-absorbing material with $n=1.45$. The transmission spectra in Figure~S2 are based on the structures in Figure~\ref{fig1}b, where the light source is set as a normal incidence plane wave. In Figure~\ref{fig3}, the complex plane transmittance was obtained by placing a point monitor 400 nm above the substrate and recording the real and imaginary components of the electric field as a function of frequency. For Figure~\ref{fig6}b, the absorbance is evaluated by the method "Power absorbed advanced" in the object library of the software. Note that we exclude the contribution of nanoparticles to the absorption by the spatial filter based on refractive index. The corresponding script is provided in the tutorial of Ansys Optics. The absorption spectra in Figure~\ref{fig6}b are evaluated by $1-T-R$, where T is the transmittance spectrum and R is the reflectance spectrum. The source is set as TE polarized wave at normal incidence. 

\subsection*{Sample Preparation}
BF$_2$ (>99\%, Lumtec) and CBP (>99.5\%, Ossila) are commercially available. The 50 wt\% CBP/BF$_2$ thin films were prepared using a chloroform solution (>99.9\%, Aldrich) at a BF$_2$ concentration of 10 mg/ml and CBP concentration of 10 mg/ml. The films are spin-coated at the rate of 1000 rpm. The film thicknesses, determined using a Dektak profilometer, were typically in the range of 100 nm. The complex refractive index of the sample was measured using a M-2000 Ellipsometer.

\subsection*{Fabrication of Aluminum nanoparticle arrays}
Al nanoparticle arrays were fabricated using electron beam lithography and a lift-off process.  First, the resist (ZEP520A) was coated onto a SiO$_2$ substrate (thickness: 500 $\mu$m) and pre-baked for 3 min at 180 $^\circ\mathrm{C}$. The resist was nanopatterned by electron beam lithography (F7000s-KYT01, Advantest) followed by development with ZED-N50. Then a thin Al film (thickness = 30 nm) was grown on the pre-patterned resist on the substrate using electron beam deposition. Finally, a lift-off process was performed with a solvent (ST-120) to remove the excess of Al on the resist.

\subsection*{Fourier Microscopy Measurements} 
The dispersion measurements of the SLRS modes were obtained using a Fourier microscope in transmission mode. The sample was illuminated through a 40× objective (Nikon CFI S Plan Fluor ELWD, NA 0.6) and the transmission was collected with a 60× objective (Nikon CFI S Plan Fluor ELWD, NA 0.7). A spectrometer (Princeton Instruments SP2300) connected to a camera (Princeton Instruments ProEM:512) allowed the mapping of the dispersion as a function of energy and angle. Extinction measurements were referenced to the extinction of air. For the PL measurements, we use the continuous-wave laser (C-WAVE) at the excitation wavelength of 532 nm. The power density was 18 mW/mm$^2$. All the measurements were performed at room temperature.

\subsection*{Coupled Oscillator Model}

The surface lattice resonance relevant in this study, with energy $E_\mathrm{SLR}$, arise from the enhanced radiative coupling of the localized resonances (LRs) in the individual nanoparticles through Rayleigh anomalies (RAs) with TE polarization and ($\pm1$,0) order. The RA energies are given by the grating equation,\cite{berghuis2020light}
\begin{align}
\label{Eq:RA}
E_\mathrm{R\pm}=\frac{c}{n_\mathrm{eff}}(\pm k_\parallel+\frac{2\pi}{\mathrm{P}}) \;,
\end{align}
where $c$ is the speed of light in vacuum, $n_\mathrm{eff}$ is the effective refractive index of the medium in which the mode propagates, and P is the period of the array.

We describe the coupling between the LRs and the RAs by the following simplified $3\times3$ Hamiltonian \cite{rodriguez2013surface}.
\begin{align}
\label{Eq:H_2}
H=\begin{pmatrix}
E_\mathrm{LR}-i\gamma_\mathrm{LR} & \Omega_\mathrm{L+} &\Omega_\mathrm{L-}\\
\Omega_\mathrm{L+} & E_\mathrm{R+}-i\gamma_\mathrm{R+}&\Omega_\pm\\ \Omega_\mathrm{L-}&\Omega_\pm&E_\mathrm{R-}-i\gamma_\mathrm{R-}
\end{pmatrix}
\;.
\end{align}
The diagonal terms in the Hamiltonian are the energies of the LRs and RAs associated with the ($\pm1$,0) diffraction orders. We fit the SLRs with the eigenstates of Eq.~\ref{Eq:H_2} using the following parameters: $E_\mathrm{LR}= 2.2$ eV, $\gamma_\mathrm{LR}=0.1$ eV, $\gamma_\mathrm{R+}=\gamma_\mathrm{R-}=2$ meV, $\Omega_\mathrm{L+}=\Omega_\mathrm{L-}=0.18$ eV and $\Omega_\mathrm{\pm}=0$. The eigenstates of Eq.~\ref{Eq:H_2} become $E_\mathrm{SLR}$ in Eq.~\ref{Eq:H}. In addition, We assume that the coupling strengths ($g$) between the SLR modes and the excitons are identical, and that their photonic dissipation rates ($\gamma_c$) are also the same.

\subsection*{TCSPC Measurements}
Time-resolved PL studies  were  performed  using a time-correlated  single  photon-counting (TCSPC)  system  in  an Edinburgh  (FLS980)  fluorometer. For the measurements of the microsecond dynamics, we apply the second harmonic of a Nd:YAG laser (Continuum, Surelite) at 532 nm as the pump pulse, with a duration of 15 ns and a repetition rate of 10 Hz. For the measurements of the microsecond dynamics, we use a picosecond diode laser (EPL-510) as the excitation source at 510 nm. All the measurements were performed at room temperature.

\begin{acknowledgement}

This project has received funding from the European Innovation Council through the project SCOLED (Grant Agreement Number 101098813). Views and opinions expressed are however those of the author(s) only and do not necessarily reflect those of the European Union or the European Innovation Council and SMEs Executive Agency (EISMEA). Neither the European Union nor the granting authority can be held responsible for them. Y.-C.W. acknowledges support from the National Science and Technology Council (NSTC) through the postdoctoral research abroad program. Y.-C.W. thank the contribution of Goudarzi Masoumeh for measuring ellipsometry and profilometry.  K.S.D has also received funding from the European Research Council (ERC) under the European Union’s Horizon 2020 research and innovation programme (grant agreement No. [948260]). S.M. acknowledges the JSPS bilateral program (JPJSBP120239921), Japan. P.-T.C acknowledges support from NCTS (NSTC 113-2639-M-002-001 -ASP).

\end{acknowledgement}

\begin{suppinfo}

The Supporting Information is available free of charge.

Ellipsometry measurement (S1); Simulated extinction spectra  (S2); SEM images (S3); Angle-resolved optical extinction measurements with TM excitation (S4); Fitting curves of the coupled oscillator model (S5); Angle-resolved optical PL measurements under TM polarization (S6); Averaged PL spectra under TE and TM polarization (S7); Nanosecond time-resolved PL (S8); Power dependent microsecond time-resolved PL (S9); Power dependent steady-state PL intensity (S10); Microsecond time-resolved PL of the substrate (S11); Fitting results of Microsecond time-resolved PL (S12)
\end{suppinfo}


\bibliography{achemso-demo}

\providecommand{\latin}[1]{#1}
\makeatletter
\providecommand{\doi}
  {\begingroup\let\do\@makeother\dospecials
  \catcode`\{=1 \catcode`\}=2 \doi@aux}
\providecommand{\doi@aux}[1]{\endgroup\texttt{#1}}
\makeatother
\providecommand*\mcitethebibliography{\thebibliography}
\csname @ifundefined\endcsname{endmcitethebibliography}  {\let\endmcitethebibliography\endthebibliography}{}
\begin{mcitethebibliography}{48}
\providecommand*\natexlab[1]{#1}
\providecommand*\mciteSetBstSublistMode[1]{}
\providecommand*\mciteSetBstMaxWidthForm[2]{}
\providecommand*\mciteBstWouldAddEndPuncttrue
  {\def\EndOfBibitem{\unskip.}}
\providecommand*\mciteBstWouldAddEndPunctfalse
  {\let\EndOfBibitem\relax}
\providecommand*\mciteSetBstMidEndSepPunct[3]{}
\providecommand*\mciteSetBstSublistLabelBeginEnd[3]{}
\providecommand*\EndOfBibitem{}
\mciteSetBstSublistMode{f}
\mciteSetBstMaxWidthForm{subitem}{(\alph{mcitesubitemcount})}
\mciteSetBstSublistLabelBeginEnd
  {\mcitemaxwidthsubitemform\space}
  {\relax}
  {\relax}

\bibitem[Uoyama \latin{et~al.}(2012)Uoyama, Goushi, Shizu, Nomura, and Adachi]{uoyama2012highly}
Uoyama,~H.; Goushi,~K.; Shizu,~K.; Nomura,~H.; Adachi,~C. Highly efficient organic light-emitting diodes from delayed fluorescence. \emph{Nature} \textbf{2012}, \emph{492}, 234--238\relax
\mciteBstWouldAddEndPuncttrue
\mciteSetBstMidEndSepPunct{\mcitedefaultmidpunct}
{\mcitedefaultendpunct}{\mcitedefaultseppunct}\relax
\EndOfBibitem
\bibitem[Wong and Zysman-Colman(2017)Wong, and Zysman-Colman]{wong2017purely}
Wong,~M.~Y.; Zysman-Colman,~E. Purely organic thermally activated delayed fluorescence materials for organic light-emitting diodes. \emph{Adv. Mater.} \textbf{2017}, \emph{29}, 1605444\relax
\mciteBstWouldAddEndPuncttrue
\mciteSetBstMidEndSepPunct{\mcitedefaultmidpunct}
{\mcitedefaultendpunct}{\mcitedefaultseppunct}\relax
\EndOfBibitem
\bibitem[Chou and Chi(2007)Chou, and Chi]{chou2007phosphorescent}
Chou,~P.-T.; Chi,~Y. Phosphorescent dyes for organic light-emitting diodes. \emph{Chem. Eur. J.} \textbf{2007}, \emph{13}, 380--395\relax
\mciteBstWouldAddEndPuncttrue
\mciteSetBstMidEndSepPunct{\mcitedefaultmidpunct}
{\mcitedefaultendpunct}{\mcitedefaultseppunct}\relax
\EndOfBibitem
\bibitem[Yang \latin{et~al.}(2015)Yang, Zhou, and Wong]{yang2015functionalization}
Yang,~X.; Zhou,~G.; Wong,~W.-Y. Functionalization of phosphorescent emitters and their host materials by main-group elements for phosphorescent organic light-emitting devices. \emph{Chem. Soc. Rev.} \textbf{2015}, \emph{44}, 8484--8575\relax
\mciteBstWouldAddEndPuncttrue
\mciteSetBstMidEndSepPunct{\mcitedefaultmidpunct}
{\mcitedefaultendpunct}{\mcitedefaultseppunct}\relax
\EndOfBibitem
\bibitem[Etherington \latin{et~al.}(2016)Etherington, Gibson, Higginbotham, Penfold, and Monkman]{etherington2016revealing}
Etherington,~M.~K.; Gibson,~J.; Higginbotham,~H.~F.; Penfold,~T.~J.; Monkman,~A.~P. Revealing the spin--vibronic coupling mechanism of thermally activated delayed fluorescence. \emph{Nat. Commun.} \textbf{2016}, \emph{7}, 13680\relax
\mciteBstWouldAddEndPuncttrue
\mciteSetBstMidEndSepPunct{\mcitedefaultmidpunct}
{\mcitedefaultendpunct}{\mcitedefaultseppunct}\relax
\EndOfBibitem
\bibitem[Jiang \latin{et~al.}(2023)Jiang, Tao, and Wong]{jiang2023recent}
Jiang,~H.; Tao,~P.; Wong,~W.-Y. Recent advances in triplet--triplet annihilation-based materials and their applications in electroluminescence. \emph{ACS Mater. Lett.} \textbf{2023}, \emph{5}, 822--845\relax
\mciteBstWouldAddEndPuncttrue
\mciteSetBstMidEndSepPunct{\mcitedefaultmidpunct}
{\mcitedefaultendpunct}{\mcitedefaultseppunct}\relax
\EndOfBibitem
\bibitem[Gao \latin{et~al.}(2021)Gao, Wong, Qin, Lo, Namdas, Dong, and Hu]{gao2021application}
Gao,~C.; Wong,~W.~W.; Qin,~Z.; Lo,~S.-C.; Namdas,~E.~B.; Dong,~H.; Hu,~W. Application of triplet--triplet annihilation upconversion in organic optoelectronic devices: advances and perspectives. \emph{Adv. Mater.} \textbf{2021}, \emph{33}, 2100704\relax
\mciteBstWouldAddEndPuncttrue
\mciteSetBstMidEndSepPunct{\mcitedefaultmidpunct}
{\mcitedefaultendpunct}{\mcitedefaultseppunct}\relax
\EndOfBibitem
\bibitem[Xiao \latin{et~al.}(2022)Xiao, Qiao, Lin, Chen, Guo, Lu, Wang, and Ma]{xiao2022situ}
Xiao,~S.; Qiao,~X.; Lin,~C.; Chen,~L.; Guo,~R.; Lu,~P.; Wang,~L.; Ma,~D. In Situ Quantifying the Physical Parameters Determining the Efficiency of OLEDs Relying on Triplet--Triplet Annihilation Up-Conversion. \emph{Adv. Opt. Mater.} \textbf{2022}, \emph{10}, 2102333\relax
\mciteBstWouldAddEndPuncttrue
\mciteSetBstMidEndSepPunct{\mcitedefaultmidpunct}
{\mcitedefaultendpunct}{\mcitedefaultseppunct}\relax
\EndOfBibitem
\bibitem[Song \latin{et~al.}(2023)Song, Li, Yu, Zhang, and He]{song2023regulating}
Song,~Y.; Li,~Y.; Yu,~R.; Zhang,~K.; He,~L. Regulating the Distance Between Donor and Acceptor in Space-Confined Through-Space Charge Transfer Emitters for Fast Radiative Decays and High-Efficiency Organic Light-Emitting Diodes With Low Efficiency Roll-Offs. \emph{Adv. Opt. Mater.} \textbf{2023}, \emph{11}, 2300432\relax
\mciteBstWouldAddEndPuncttrue
\mciteSetBstMidEndSepPunct{\mcitedefaultmidpunct}
{\mcitedefaultendpunct}{\mcitedefaultseppunct}\relax
\EndOfBibitem
\bibitem[Kim \latin{et~al.}(2020)Kim, Park, Chan, Tanaka, Tsuchiya, Nakanotani, and Adachi]{kim2020nanosecond}
Kim,~J.~U.; Park,~I.~S.; Chan,~C.-Y.; Tanaka,~M.; Tsuchiya,~Y.; Nakanotani,~H.; Adachi,~C. Nanosecond-time-scale delayed fluorescence molecule for deep-blue OLEDs with small efficiency rolloff. \emph{Nat. Commun.} \textbf{2020}, \emph{11}, 1765\relax
\mciteBstWouldAddEndPuncttrue
\mciteSetBstMidEndSepPunct{\mcitedefaultmidpunct}
{\mcitedefaultendpunct}{\mcitedefaultseppunct}\relax
\EndOfBibitem
\bibitem[Reineke \latin{et~al.}(2007)Reineke, Walzer, and Leo]{reineke2007triplet}
Reineke,~S.; Walzer,~K.; Leo,~K. Triplet-exciton quenching in organic phosphorescent light-emitting diodes with Ir-based emitters. \emph{Phys. Rev. B} \textbf{2007}, \emph{75}, 125328\relax
\mciteBstWouldAddEndPuncttrue
\mciteSetBstMidEndSepPunct{\mcitedefaultmidpunct}
{\mcitedefaultendpunct}{\mcitedefaultseppunct}\relax
\EndOfBibitem
\bibitem[Zhang and Forrest(2012)Zhang, and Forrest]{zhang2012triplets}
Zhang,~Y.; Forrest,~S.~R. Triplets Contribute to Both an Increase and Loss in Fluorescent Yield in Organic Light Emitting Diodes. \emph{Phys. Rev. Lett.} \textbf{2012}, \emph{108}, 267404\relax
\mciteBstWouldAddEndPuncttrue
\mciteSetBstMidEndSepPunct{\mcitedefaultmidpunct}
{\mcitedefaultendpunct}{\mcitedefaultseppunct}\relax
\EndOfBibitem
\bibitem[Virgili \latin{et~al.}(2011)Virgili, Coles, Adawi, Clark, Michetti, Rajendran, Brida, Polli, Cerullo, and Lidzey]{virgili2011ultrafast}
Virgili,~T.; Coles,~D.; Adawi,~A.; Clark,~C.; Michetti,~P.; Rajendran,~S.; Brida,~D.; Polli,~D.; Cerullo,~G.; Lidzey,~D. Ultrafast polariton relaxation dynamics in an organic semiconductor microcavity. \emph{Phys. Rev. B} \textbf{2011}, \emph{83}, 245309\relax
\mciteBstWouldAddEndPuncttrue
\mciteSetBstMidEndSepPunct{\mcitedefaultmidpunct}
{\mcitedefaultendpunct}{\mcitedefaultseppunct}\relax
\EndOfBibitem
\bibitem[Hutchison \latin{et~al.}(2012)Hutchison, Schwartz, Genet, Devaux, and Ebbesen]{hutchison2012modifying}
Hutchison,~J.~A.; Schwartz,~T.; Genet,~C.; Devaux,~E.; Ebbesen,~T.~W. Modifying chemical landscapes by coupling to vacuum fields. \emph{Angew. Chem. Int. Ed.} \textbf{2012}, \emph{51}, 1592--1596\relax
\mciteBstWouldAddEndPuncttrue
\mciteSetBstMidEndSepPunct{\mcitedefaultmidpunct}
{\mcitedefaultendpunct}{\mcitedefaultseppunct}\relax
\EndOfBibitem
\bibitem[Schwartz \latin{et~al.}(2013)Schwartz, Hutchison, L{\'e}onard, Genet, Haacke, and Ebbesen]{schwartz2013polariton}
Schwartz,~T.; Hutchison,~J.~A.; L{\'e}onard,~J.; Genet,~C.; Haacke,~S.; Ebbesen,~T.~W. Polariton dynamics under strong light--molecule coupling. \emph{ChemPhysChem} \textbf{2013}, \emph{14}, 125--131\relax
\mciteBstWouldAddEndPuncttrue
\mciteSetBstMidEndSepPunct{\mcitedefaultmidpunct}
{\mcitedefaultendpunct}{\mcitedefaultseppunct}\relax
\EndOfBibitem
\bibitem[Wang \latin{et~al.}(2014)Wang, Chervy, George, Hutchison, Genet, and Ebbesen]{wang2014quantum}
Wang,~S.; Chervy,~T.; George,~J.; Hutchison,~J.~A.; Genet,~C.; Ebbesen,~T.~W. Quantum yield of polariton emission from hybrid light-matter states. \emph{J. Phys. Chem. Lett.} \textbf{2014}, \emph{5}, 1433--1439\relax
\mciteBstWouldAddEndPuncttrue
\mciteSetBstMidEndSepPunct{\mcitedefaultmidpunct}
{\mcitedefaultendpunct}{\mcitedefaultseppunct}\relax
\EndOfBibitem
\bibitem[Balasubrahmaniyam \latin{et~al.}(2023)Balasubrahmaniyam, Simkhovich, Golombek, Sandik, Ankonina, and Schwartz]{balasubrahmaniyam2023enhanced}
Balasubrahmaniyam,~M.; Simkhovich,~A.; Golombek,~A.; Sandik,~G.; Ankonina,~G.; Schwartz,~T. From enhanced diffusion to ultrafast ballistic motion of hybrid light--matter excitations. \emph{Nat. Mater.} \textbf{2023}, \emph{22}, 338--344\relax
\mciteBstWouldAddEndPuncttrue
\mciteSetBstMidEndSepPunct{\mcitedefaultmidpunct}
{\mcitedefaultendpunct}{\mcitedefaultseppunct}\relax
\EndOfBibitem
\bibitem[Tibben \latin{et~al.}(2023)Tibben, Bonin, Cho, Lakhwani, Hutchison, and G{\'o}mez]{tibben2023molecular}
Tibben,~D.~J.; Bonin,~G.~O.; Cho,~I.; Lakhwani,~G.; Hutchison,~J.; G{\'o}mez,~D.~E. Molecular energy transfer under the strong light--matter interaction regime. \emph{Chem. Rev.} \textbf{2023}, \emph{123}, 8044--8068\relax
\mciteBstWouldAddEndPuncttrue
\mciteSetBstMidEndSepPunct{\mcitedefaultmidpunct}
{\mcitedefaultendpunct}{\mcitedefaultseppunct}\relax
\EndOfBibitem
\bibitem[Zeng \latin{et~al.}(2023)Zeng, P{\'e}rez-S{\'a}nchez, Eckdahl, Liu, Chang, Weiss, Kalow, Yuen-Zhou, and Stern]{zeng2023control}
Zeng,~H.; P{\'e}rez-S{\'a}nchez,~J.~B.; Eckdahl,~C.~T.; Liu,~P.; Chang,~W.~J.; Weiss,~E.~A.; Kalow,~J.~A.; Yuen-Zhou,~J.; Stern,~N.~P. Control of photoswitching kinetics with strong light--matter coupling in a cavity. \emph{J. Am. Chem. Soc.} \textbf{2023}, \emph{145}, 19655--19661\relax
\mciteBstWouldAddEndPuncttrue
\mciteSetBstMidEndSepPunct{\mcitedefaultmidpunct}
{\mcitedefaultendpunct}{\mcitedefaultseppunct}\relax
\EndOfBibitem
\bibitem[Nikolis \latin{et~al.}(2019)Nikolis, Mischok, Siegmund, Kublitski, Jia, Benduhn, H{\"o}rmann, Neher, Gather, Spoltore, \latin{et~al.} others]{nikolis2019strong}
Nikolis,~V.~C.; Mischok,~A.; Siegmund,~B.; Kublitski,~J.; Jia,~X.; Benduhn,~J.; H{\"o}rmann,~U.; Neher,~D.; Gather,~M.~C.; Spoltore,~D.; others Strong light-matter coupling for reduced photon energy losses in organic photovoltaics. \emph{Nat. Commun.} \textbf{2019}, \emph{10}, 3706\relax
\mciteBstWouldAddEndPuncttrue
\mciteSetBstMidEndSepPunct{\mcitedefaultmidpunct}
{\mcitedefaultendpunct}{\mcitedefaultseppunct}\relax
\EndOfBibitem
\bibitem[Tang \latin{et~al.}(2024)Tang, Stuart, van~der Laan, and Lakhwani]{tang2024strong}
Tang,~Y.; Stuart,~A.~N.; van~der Laan,~T.; Lakhwani,~G. Strong Light--Matter Coupling Leads to a Longer Charge Carrier Lifetime in Cavity Organic Solar Cells. \emph{ACS Photonics} \textbf{2024}, \emph{11}, 1627--1637\relax
\mciteBstWouldAddEndPuncttrue
\mciteSetBstMidEndSepPunct{\mcitedefaultmidpunct}
{\mcitedefaultendpunct}{\mcitedefaultseppunct}\relax
\EndOfBibitem
\bibitem[de~Jong \latin{et~al.}(2024)de~Jong, Berghuis, Abdelkhalik, van~der Pol, Wienk, Janssen, and G{\'o}mez~Rivas]{de2024enhancement}
de~Jong,~L.~M.; Berghuis,~A.~M.; Abdelkhalik,~M.~S.; van~der Pol,~T.~P.; Wienk,~M.~M.; Janssen,~R.~A.; G{\'o}mez~Rivas,~J. Enhancement of the internal quantum efficiency in strongly coupled p3ht-c60 organic photovoltaic cells using fabry--perot cavities with varied cavity confinement. \emph{Nanophotonics} \textbf{2024}, \emph{13}, 2531--2540\relax
\mciteBstWouldAddEndPuncttrue
\mciteSetBstMidEndSepPunct{\mcitedefaultmidpunct}
{\mcitedefaultendpunct}{\mcitedefaultseppunct}\relax
\EndOfBibitem
\bibitem[Mischok \latin{et~al.}(2023)Mischok, Hillebrandt, Kwon, and Gather]{mischok2023highly}
Mischok,~A.; Hillebrandt,~S.; Kwon,~S.; Gather,~M.~C. Highly efficient polaritonic light-emitting diodes with angle-independent narrowband emission. \emph{Nat. Photonics} \textbf{2023}, \emph{17}, 393--400\relax
\mciteBstWouldAddEndPuncttrue
\mciteSetBstMidEndSepPunct{\mcitedefaultmidpunct}
{\mcitedefaultendpunct}{\mcitedefaultseppunct}\relax
\EndOfBibitem
\bibitem[Witt \latin{et~al.}(2024)Witt, Mischok, Tenopala~Carmona, Hillebrandt, Butscher, and Gather]{witt2024high}
Witt,~J.; Mischok,~A.; Tenopala~Carmona,~F.; Hillebrandt,~S.; Butscher,~J.~F.; Gather,~M.~C. High-brightness blue polariton organic light-emitting diodes. \emph{ACS photonics} \textbf{2024}, \emph{11}, 1844--1850\relax
\mciteBstWouldAddEndPuncttrue
\mciteSetBstMidEndSepPunct{\mcitedefaultmidpunct}
{\mcitedefaultendpunct}{\mcitedefaultseppunct}\relax
\EndOfBibitem
\bibitem[Ebbesen(2016)]{ebbesen2016hybrid}
Ebbesen,~T.~W. Hybrid light--matter states in a molecular and material science perspective. \emph{Acc. Chem. Res.} \textbf{2016}, \emph{49}, 2403--2412\relax
\mciteBstWouldAddEndPuncttrue
\mciteSetBstMidEndSepPunct{\mcitedefaultmidpunct}
{\mcitedefaultendpunct}{\mcitedefaultseppunct}\relax
\EndOfBibitem
\bibitem[Bhuyan \latin{et~al.}(2023)Bhuyan, Mony, Kotov, Castellanos, Gómez~Rivas, Shegai, and B{\"o}rjesson]{bhuyan2023rise}
Bhuyan,~R.; Mony,~J.; Kotov,~O.; Castellanos,~G.~W.; Gómez~Rivas,~J.; Shegai,~T.~O.; B{\"o}rjesson,~K. The rise and current status of polaritonic photochemistry and photophysics. \emph{Chem. Rev.} \textbf{2023}, \emph{123}, 10877--10919\relax
\mciteBstWouldAddEndPuncttrue
\mciteSetBstMidEndSepPunct{\mcitedefaultmidpunct}
{\mcitedefaultendpunct}{\mcitedefaultseppunct}\relax
\EndOfBibitem
\bibitem[Simpkins \latin{et~al.}(2023)Simpkins, Dunkelberger, and Vurgaftman]{simpkins2023control}
Simpkins,~B.~S.; Dunkelberger,~A.~D.; Vurgaftman,~I. Control, modulation, and analytical descriptions of vibrational strong coupling. \emph{Chem. Rev.} \textbf{2023}, \emph{123}, 5020--5048\relax
\mciteBstWouldAddEndPuncttrue
\mciteSetBstMidEndSepPunct{\mcitedefaultmidpunct}
{\mcitedefaultendpunct}{\mcitedefaultseppunct}\relax
\EndOfBibitem
\bibitem[Eizner \latin{et~al.}(2019)Eizner, Mart{\'\i}nez-Mart{\'\i}nez, Yuen-Zhou, and K{\'e}na-Cohen]{eizner2019inverting}
Eizner,~E.; Mart{\'\i}nez-Mart{\'\i}nez,~L.~A.; Yuen-Zhou,~J.; K{\'e}na-Cohen,~S. Inverting singlet and triplet excited states using strong light-matter coupling. \emph{Sci. Adv.} \textbf{2019}, \emph{5}, eaax4482\relax
\mciteBstWouldAddEndPuncttrue
\mciteSetBstMidEndSepPunct{\mcitedefaultmidpunct}
{\mcitedefaultendpunct}{\mcitedefaultseppunct}\relax
\EndOfBibitem
\bibitem[Cho \latin{et~al.}(2023)Cho, Kendrick, Stuart, Ramkissoon, Ghiggino, Wong, and Lakhwani]{cho2023multi}
Cho,~I.; Kendrick,~W.~J.; Stuart,~A.~N.; Ramkissoon,~P.; Ghiggino,~K.~P.; Wong,~W.~W.; Lakhwani,~G. Multi-resonance TADF in optical cavities: suppressing excimer emission through efficient energy transfer to the lower polariton states. \emph{J. Mater. Chem. C} \textbf{2023}, \emph{11}, 14448--14455\relax
\mciteBstWouldAddEndPuncttrue
\mciteSetBstMidEndSepPunct{\mcitedefaultmidpunct}
{\mcitedefaultendpunct}{\mcitedefaultseppunct}\relax
\EndOfBibitem
\bibitem[Abdelmagid \latin{et~al.}(2024)Abdelmagid, Qureshi, Papachatzakis, Siltanen, Kumar, Ashokan, Salman, Luoma, and Daskalakis]{abdelmagid2024identifying}
Abdelmagid,~A.~G.; Qureshi,~H.~A.; Papachatzakis,~M.~A.; Siltanen,~O.; Kumar,~M.; Ashokan,~A.; Salman,~S.; Luoma,~K.; Daskalakis,~K.~S. Identifying the origin of delayed electroluminescence in a polariton organic light-emitting diode. \emph{Nanophotonics} \textbf{2024}, \emph{13}, 2565--2573\relax
\mciteBstWouldAddEndPuncttrue
\mciteSetBstMidEndSepPunct{\mcitedefaultmidpunct}
{\mcitedefaultendpunct}{\mcitedefaultseppunct}\relax
\EndOfBibitem
\bibitem[Ishii \latin{et~al.}(2024)Ishii, P{\'e}rez-S{\'a}nchez, Yuen-Zhou, Adachi, Hatakeyama, and K{\'e}na-Cohen]{ishii2024modified}
Ishii,~T.; P{\'e}rez-S{\'a}nchez,~J.~B.; Yuen-Zhou,~J.; Adachi,~C.; Hatakeyama,~T.; K{\'e}na-Cohen,~S. Modified Prompt and Delayed Kinetics in a Strongly Coupled Organic Microcavity Containing a Multiresonance TADF Emitter. \emph{ACS Photonics} \textbf{2024}, \emph{11}, 3998--4007\relax
\mciteBstWouldAddEndPuncttrue
\mciteSetBstMidEndSepPunct{\mcitedefaultmidpunct}
{\mcitedefaultendpunct}{\mcitedefaultseppunct}\relax
\EndOfBibitem
\bibitem[Polak \latin{et~al.}(2020)Polak, Jayaprakash, Lyons, Martínez-Martínez, Leventis, Fallon, Coulthard, Bossanyi, Georgiou, Petty{,}~II, Anthony, Bronstein, Yuen-Zhou, Tartakovskii, Clark, and Musser]{polak2020manipulating}
Polak,~D. \latin{et~al.}  Manipulating molecules with strong coupling: harvesting triplet excitons in organic exciton microcavities. \emph{Chem. Sci.} \textbf{2020}, \emph{11}, 343--354\relax
\mciteBstWouldAddEndPuncttrue
\mciteSetBstMidEndSepPunct{\mcitedefaultmidpunct}
{\mcitedefaultendpunct}{\mcitedefaultseppunct}\relax
\EndOfBibitem
\bibitem[Berghuis \latin{et~al.}(2019)Berghuis, Halpin, Le-Van, Ramezani, Wang, Murai, and G{\'o}mez~Rivas]{berghuis2019enhanced}
Berghuis,~A.~M.; Halpin,~A.; Le-Van,~Q.; Ramezani,~M.; Wang,~S.; Murai,~S.; G{\'o}mez~Rivas,~J. Enhanced delayed fluorescence in tetracene crystals by strong light-matter coupling. \emph{Adv. Funct. Mater.} \textbf{2019}, \emph{29}, 1901317\relax
\mciteBstWouldAddEndPuncttrue
\mciteSetBstMidEndSepPunct{\mcitedefaultmidpunct}
{\mcitedefaultendpunct}{\mcitedefaultseppunct}\relax
\EndOfBibitem
\bibitem[Kim \latin{et~al.}(2018)Kim, D’Al{\'e}o, Chen, Sandanayaka, Yao, Zhao, Komino, Zaborova, Canard, Tsuchiya, Choi, Wu, Fages, Br{\'e}das, Ribierre, and Adachi]{kim2018high}
Kim,~D.-H. \latin{et~al.}  High-efficiency electroluminescence and amplified spontaneous emission from a thermally activated delayed fluorescent near-infrared emitter. \emph{Nat. Photonics} \textbf{2018}, \emph{12}, 98--104\relax
\mciteBstWouldAddEndPuncttrue
\mciteSetBstMidEndSepPunct{\mcitedefaultmidpunct}
{\mcitedefaultendpunct}{\mcitedefaultseppunct}\relax
\EndOfBibitem
\bibitem[Adachi \latin{et~al.}(2001)Adachi, Kwong, Djurovich, Adamovich, Baldo, Thompson, and Forrest]{adachi2001endothermic}
Adachi,~C.; Kwong,~R.~C.; Djurovich,~P.; Adamovich,~V.; Baldo,~M.~A.; Thompson,~M.~E.; Forrest,~S.~R. Endothermic energy transfer: A mechanism for generating very efficient high-energy phosphorescent emission in organic materials. \emph{Appl. Phys. Lett.} \textbf{2001}, \emph{79}, 2082--2084\relax
\mciteBstWouldAddEndPuncttrue
\mciteSetBstMidEndSepPunct{\mcitedefaultmidpunct}
{\mcitedefaultendpunct}{\mcitedefaultseppunct}\relax
\EndOfBibitem
\bibitem[Dias \latin{et~al.}(2013)Dias, Bourdakos, Jankus, Moss, Kamtekar, Bhalla, Santos, Bryce, and Monkman]{dias2013triplet}
Dias,~F.~B.; Bourdakos,~K.~N.; Jankus,~V.; Moss,~K.~C.; Kamtekar,~K.~T.; Bhalla,~V.; Santos,~J.; Bryce,~M.~R.; Monkman,~A.~P. Triplet harvesting with 100\% efficiency by way of thermally activated delayed fluorescence in charge transfer OLED emitters. \emph{Adv. Mater.} \textbf{2013}, \emph{25}, 3707--3714\relax
\mciteBstWouldAddEndPuncttrue
\mciteSetBstMidEndSepPunct{\mcitedefaultmidpunct}
{\mcitedefaultendpunct}{\mcitedefaultseppunct}\relax
\EndOfBibitem
\bibitem[Lepp{\"a}l{\"a} \latin{et~al.}(2024)Lepp{\"a}l{\"a}, Abdelmagid, Qureshi, Daskalakis, and Luoma]{leppala2024linear}
Lepp{\"a}l{\"a},~T.; Abdelmagid,~A.~G.; Qureshi,~H.~A.; Daskalakis,~K.~S.; Luoma,~K. Linear optical properties of organic microcavity polaritons with non-Markovian quantum state diffusion. \emph{Nanophotonics} \textbf{2024}, \emph{13}, 2479--2490\relax
\mciteBstWouldAddEndPuncttrue
\mciteSetBstMidEndSepPunct{\mcitedefaultmidpunct}
{\mcitedefaultendpunct}{\mcitedefaultseppunct}\relax
\EndOfBibitem
\bibitem[{Lumerical, Inc.}()]{FDTD}
{Lumerical, Inc.} FDTD Solutions\relax
\mciteBstWouldAddEndPuncttrue
\mciteSetBstMidEndSepPunct{\mcitedefaultmidpunct}
{\mcitedefaultendpunct}{\mcitedefaultseppunct}\relax
\EndOfBibitem
\bibitem[Verdelli \latin{et~al.}(2024)Verdelli, Wei, Joseph, Abdelkhalik, Goudarzi, Askes, Baldi, Meijer, and Gomez~Rivas]{verdelli2024polaritonic}
Verdelli,~F.; Wei,~Y.-C.; Joseph,~K.; Abdelkhalik,~M.~S.; Goudarzi,~M.; Askes,~S.~H.; Baldi,~A.; Meijer,~E.; Gomez~Rivas,~J. Polaritonic Chemistry Enabled by Non-Local Metasurfaces. \emph{Angew. Chem. Int. Ed.} \textbf{2024}, \emph{63}, e202409528\relax
\mciteBstWouldAddEndPuncttrue
\mciteSetBstMidEndSepPunct{\mcitedefaultmidpunct}
{\mcitedefaultendpunct}{\mcitedefaultseppunct}\relax
\EndOfBibitem
\bibitem[Verdelli \latin{et~al.}(2024)Verdelli, Wei, Scheers, Abdelkhalik, Goudarzi, and G{\'o}mez~Rivas]{verdelli2024ultrastrong}
Verdelli,~F.; Wei,~Y.-C.; Scheers,~J.~M.; Abdelkhalik,~M.~S.; Goudarzi,~M.; G{\'o}mez~Rivas,~J. Ultrastrong coupling between molecular vibrations in water and surface lattice resonances. \emph{J. Chem. Phys.} \textbf{2024}, \emph{161}\relax
\mciteBstWouldAddEndPuncttrue
\mciteSetBstMidEndSepPunct{\mcitedefaultmidpunct}
{\mcitedefaultendpunct}{\mcitedefaultseppunct}\relax
\EndOfBibitem
\bibitem[Liu \latin{et~al.}(2005)Liu, Kwong, Cheung, Djuri{\v{s}}i{\'c}, Chan, and Chui]{liu2005characterization}
Liu,~Z.; Kwong,~C.; Cheung,~C.; Djuri{\v{s}}i{\'c},~A.; Chan,~Y.; Chui,~P. The characterization of the optical functions of BCP and CBP thin films by spectroscopic ellipsometry. \emph{Synth. Met.} \textbf{2005}, \emph{150}, 159--163\relax
\mciteBstWouldAddEndPuncttrue
\mciteSetBstMidEndSepPunct{\mcitedefaultmidpunct}
{\mcitedefaultendpunct}{\mcitedefaultseppunct}\relax
\EndOfBibitem
\bibitem[Morozov \latin{et~al.}(2019)Morozov, Ivanov, de~Sa~Pereira, Menelaou, Monkman, Pozina, and Kaliteevski]{morozov2019revising}
Morozov,~K.~M.; Ivanov,~K.~A.; de~Sa~Pereira,~D.; Menelaou,~C.; Monkman,~A.~P.; Pozina,~G.; Kaliteevski,~M.~A. Revising of the Purcell effect in periodic metal-dielectric structures: the role of absorption. \emph{Sci. Rep.} \textbf{2019}, \emph{9}, 9604\relax
\mciteBstWouldAddEndPuncttrue
\mciteSetBstMidEndSepPunct{\mcitedefaultmidpunct}
{\mcitedefaultendpunct}{\mcitedefaultseppunct}\relax
\EndOfBibitem
\bibitem[Savona \latin{et~al.}(1995)Savona, Andreani, Schwendimann, and Quattropani]{savona1995quantum}
Savona,~V.; Andreani,~L.; Schwendimann,~P.; Quattropani,~A. Quantum well excitons in semiconductor microcavities: Unified treatment of weak and strong coupling regimes. \emph{Solid State Commun.} \textbf{1995}, \emph{93}, 733--739\relax
\mciteBstWouldAddEndPuncttrue
\mciteSetBstMidEndSepPunct{\mcitedefaultmidpunct}
{\mcitedefaultendpunct}{\mcitedefaultseppunct}\relax
\EndOfBibitem
\bibitem[Thomas \latin{et~al.}(2020)Thomas, Tan, Fernandez, and Barnes]{thomas2020new}
Thomas,~P.~A.; Tan,~W.~J.; Fernandez,~H.~A.; Barnes,~W.~L. A new signature for strong light--matter coupling using spectroscopic ellipsometry. \emph{Nano Lett.} \textbf{2020}, \emph{20}, 6412--6419\relax
\mciteBstWouldAddEndPuncttrue
\mciteSetBstMidEndSepPunct{\mcitedefaultmidpunct}
{\mcitedefaultendpunct}{\mcitedefaultseppunct}\relax
\EndOfBibitem
\bibitem[Cheng \latin{et~al.}(2016)Cheng, Su, Choi, Gwo, Li, and Shih]{cheng2016epitaxial}
Cheng,~F.; Su,~P.-H.; Choi,~J.; Gwo,~S.; Li,~X.; Shih,~C.-K. Epitaxial growth of atomically smooth aluminum on silicon and its intrinsic optical properties. \emph{ACS Nano} \textbf{2016}, \emph{10}, 9852--9860\relax
\mciteBstWouldAddEndPuncttrue
\mciteSetBstMidEndSepPunct{\mcitedefaultmidpunct}
{\mcitedefaultendpunct}{\mcitedefaultseppunct}\relax
\EndOfBibitem
\bibitem[Berghuis \latin{et~al.}(2020)Berghuis, Serpenti, Ramezani, Wang, and G{\'o}mez~Rivas]{berghuis2020light}
Berghuis,~A.~M.; Serpenti,~V.; Ramezani,~M.; Wang,~S.; G{\'o}mez~Rivas,~J. Light-matter coupling strength controlled by the orientation of organic crystals in plasmonic cavities. \emph{J. Phys. Chem. C} \textbf{2020}, \emph{124}, 12030--12038\relax
\mciteBstWouldAddEndPuncttrue
\mciteSetBstMidEndSepPunct{\mcitedefaultmidpunct}
{\mcitedefaultendpunct}{\mcitedefaultseppunct}\relax
\EndOfBibitem
\bibitem[Rodriguez and Rivas(2013)Rodriguez, and Rivas]{rodriguez2013surface}
Rodriguez,~S.; Rivas,~J.~G. Surface lattice resonances strongly coupled to Rhodamine 6G excitons: tuning the plasmon-exciton-polariton mass and composition. \emph{Opt. Express} \textbf{2013}, \emph{21}, 27411--27421\relax
\mciteBstWouldAddEndPuncttrue
\mciteSetBstMidEndSepPunct{\mcitedefaultmidpunct}
{\mcitedefaultendpunct}{\mcitedefaultseppunct}\relax
\EndOfBibitem
\end{mcitethebibliography}

\end{document}